\documentclass[aip,amsmath,amssymb,reprint]{revtex4-1}
\usepackage{graphicx}
\usepackage{dcolumn}
\usepackage{bm}
\usepackage{upgreek}
\usepackage{multirow}
\usepackage[utf8]{inputenc}
\usepackage[T1]{fontenc}
\usepackage{mathptmx}
\usepackage{hyperref}
\usepackage{color}
\hypersetup{colorlinks,citecolor=blue, filecolor=blue ,linkcolor=blue , urlcolor=blue, pdftex}

\usepackage[labeled,resetlabels]{multibib}

\newcites{SM}{SupM}

\begin{document}

\title[]{Breathing modes in few-layer MoTe$_2$ activated by h-BN encapsulation}

\author{M. Grzeszczyk}
\email{magdalena.grzeszczyk@fuw.edu.pl}

\author{M. R. Molas}
\affiliation{Institute of Experimental Physics, Faculty of Physics, University of Warsaw, ul.
Pasteura 5, 02-093 Warsaw, Poland}

\author{M. Barto\v{s}}
\affiliation{Laboratoire National des Champs Magnétiques Intenses, CNRS-UGA-UPS-INSA-
EMFL, 25, avenue des Martyrs, 38042 Grenoble, France}
\affiliation{Central European Institute of Technology, Brno University of Technology,  Purky\v{n}ova 656/123, 612 00 Brno, Czech Republic}

\author{K. Nogajewski}
\affiliation{Institute of Experimental Physics, Faculty of Physics, University of Warsaw, ul.
Pasteura 5, 02-093 Warsaw, Poland}

\author{M. Potemski}
\affiliation{Institute of Experimental Physics, Faculty of Physics, University of Warsaw, ul.
Pasteura 5, 02-093 Warsaw, Poland}
\affiliation{Laboratoire National des Champs Magnétiques Intenses, CNRS-UGA-UPS-INSA-
EMFL, 25, avenue des Martyrs, 38042 Grenoble, France}

\author{A. Babiński}
\affiliation{Institute of Experimental Physics, Faculty of Physics, University of Warsaw, ul.
Pasteura 5, 02-093 Warsaw, Poland}

\begin{abstract}

The encapsulation of few-layer transition metal dichalcogenides (TMDs) in hexagonal boron nitride (h-BN) is known to improve significantly their optical and electronic properties. However, it may be expected that the h-BN encapsulation may affect also vibration properties of TMDs due to an atomically flat surface of h-BN layers. In order to study its effect on interlayer interactions in few-layer TMDs, we investigate low-energy Raman scattering spectra of bi- and trilayer MoTe$_2$. Surprisingly, three breathing modes are observed in the Raman spectra of the structures deposited on or encapsulated in h-BN as compared to a single breathing mode for the flakes deposited on a SiO$_2$/Si substrate. The shear mode is not affected by changing the MoTe$_2$ environment. The emerged structure of breathing modes is ascribed to the apparent interaction between the MoTe$_2$ layer and the bottom h-BN flake. The structure becomes visible due to a high-quality surface of the former flake. Consequently, the observed triple structure of breathing modes originates from the combination modes due to interlayer and layer--substrate interactions. Our results confirm that the h-BN encapsulation affects substantially vibration properties of layered materials.
\end{abstract}

\maketitle


Two--dimensional van der Waals layered materials, including transition metal dichalcogenides (TMDs), have recently emerged as promising candidates for many applications \cite{radisavljevic2011single,choi2013controlled,wang2012integrated}. However, due to the large surface to volume ratio, their optical and electronic properties may to a large extent be dictated by the quality of the surface on which they are deposited. First reports on pristine TMD flakes showed that linewidths of free exciton emission exhibited significant inhomogeneous broadening, which was typically attributed to the local spatial inhomogeneity of the substrate, strain, and atoms/molecules adsorbed on the surface \cite{zhou2016electronic,ajayi2017approaching,tongay2013broad,su2015effects}. Moreover, the airtight sealing of TMD layers between h-BN films reduces disorder due to an atomically flat h-BN surface and protects the TMD material from contamination and external conditions. \cite{lee2015highly,cao2015quality,ahn2016prevention} This results in substantial improvements of the emission spectra of h-BN--encapsulated TMD monolayers (1~L) as compared to non-encapsulated ones, allowing for detailed studies of numerous excitonic features \cite{cadiz2017excitonic, molas2019energy}. Improvements of TMDs electronic properties were also observed. The h-BN-encapsulated MoS$_2$ reveals ultrahigh low-temperature mobility \cite{cui2014multi} and the micrometer-scale ballistic transport is maintained at room temperature in high-quality h-BN-encapsulated graphene devices  \cite{mayorov2011micrometer}. The examples show the key importance of the substrate surface for layered materials. However, a vast majority of studies on van der Waals structures do not take into account interlayer interactions, which so far have been generally neglected \cite{lee2015resonant,froehlicher2015unified}. Exploiting interlayer interactions, especially between electrons and phonons, allows a new degree of freedom, which could be applied in devices and result in novel functionalities that have not been previously possible.

We demonstrate the effect of the substrate surface on the rigid layer modes in thin layers of molybdenum ditelluride (MoTe$_2$). The interaction between the MoTe$_2$ layer and the surrounding medium is studied by means of the low-frequency Raman scattering spectroscopy in bi- (2~L) and trilayer (3~L) MoTe$_2$. We observe additional peaks due to the breathing modes (BMs) in the Raman spectra of MoTe$_2$ layers placed on the h-BN substrate, while no effect on the shear modes (SMs) is apparent. We propose that their emergence results from a well-defined interaction between the lowest MoTe$_2$ layer and the h-BN substrate.

\begin{figure*}[t!]
    \centering
    \includegraphics[width=\linewidth]{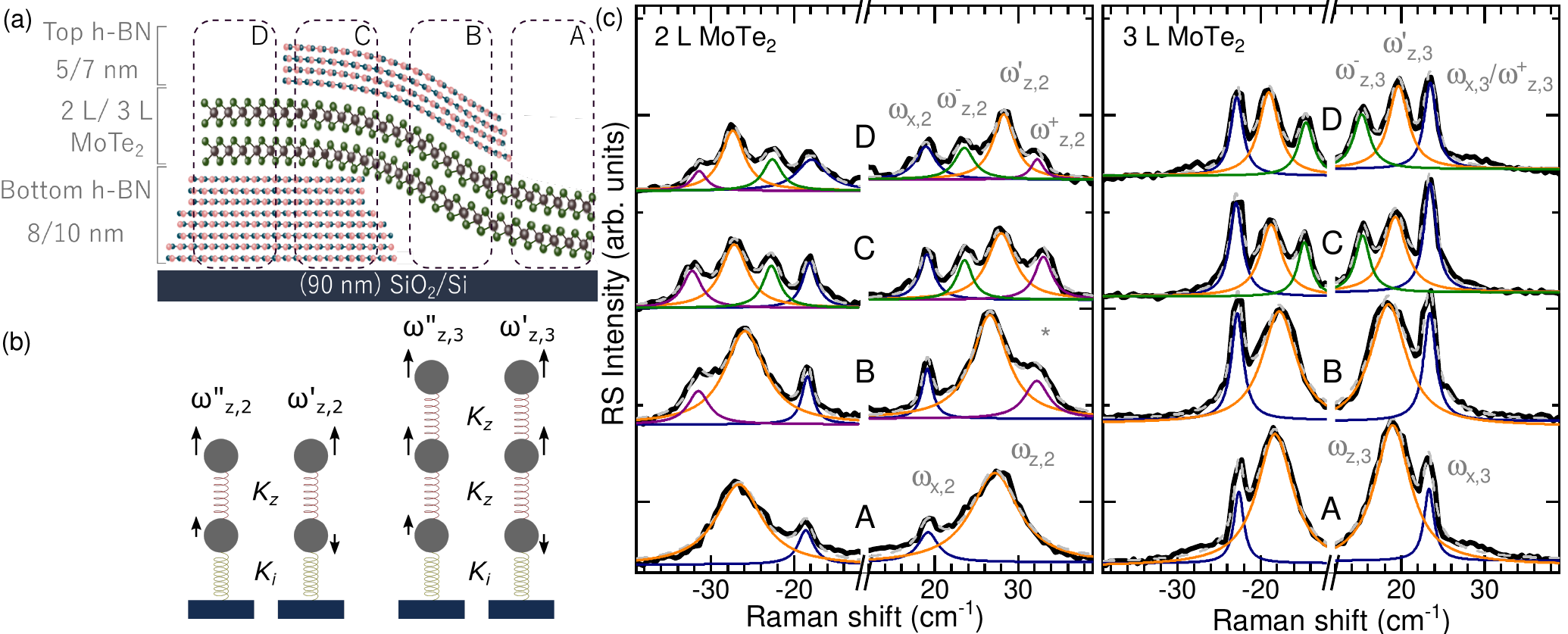}
    \caption{(a) Scheme of the investigated sample structure. The A, B, C, and D regions correspond to MoTe$_2$/SiO$_2$/Si, h-BN/MoTe$_2$/SiO$_2$/Si, \mbox{h-BN/MoTe$_2$/h-BN/SiO$_2$/Si}, and MoTe$_2$/h-BN/SiO$_2$/Si heterostructures, respectively. (b) Illustration of the atomic displacements of the modes $\omega'_{\text{z},N}$ and $\omega''_{\text{z},N}$, where index $N$ corresponds to a number of layers. Each ball represents a single layer of MoTe$_2$ and a bar depicts h-BN flakes. (c) Low-frequency Raman spectra (E$_\text{L}$=1.96~eV) of 2~L and 3~L MoTe$_2$ in the A, B, C, and D regions.}
    \label{fig:196}
    \vspace{-10pt}
\end{figure*}

The investigated samples comprised thin MoTe$_2$ and \mbox{h-BN} layers were fabricated by two-stage PDMS-based mechanical exfoliation, described in detail in Supplementary Material (SM). The 2~L (3~L) MoTe$_2$ flake was placed on SiO$_2$ (90~nm)/Si substrate, which was partially covered with a 7~nm (10~nm) thick h-BN layer. The heterostructure was partially capped with a 5~nm (8~nm) thick top h-BN flake. As a result, four regions of the samples could be identified: MoTe$_2$/SiO$_2$/Si, h-BN/MoTe$_2$/SiO$_2$/Si, h-BN/MoTe$_2$/h-BN/SiO$_2$/Si, and MoTe$_2$/h-BN/SiO$_2$/Si, which in the following are referred to as A, B, C, and D respectively. The schematic representation of the sample structure is shown in Fig. \ref{fig:196}(a).


Unpolarized low-frequency Stokes and anti-Stokes Raman scattering spectra measured on the heterostructures under \mbox{E$_\text{L}$=1.96~eV} excitation are shown in Fig. \ref{fig:196}(c). The spectra from regions A comprise two peaks, which are characteristic of atomically thin MoTe$_2$ \cite{froehlicher2015unified,grzeszczyk2016raman} or other thin TMD structures \cite{boukhicha2013anharmonic,zhao2013interlayer,zhang2013raman,kim2017excitonic}. The peaks correspond to rigid displacements of whole Te--Mo--Te layers, which are parallel (shear mode - SM) and perpendicular (breathing mode - BM) to the structure plane. Their energies are summarized in Table \ref{tab:table}. The BM softens (redshifts) while the SM hardens (blueshifts) with an increasing number of layers. \cite{froehlicher2015unified,grzeszczyk2016raman} As a result the BM in 2~L-MoTe$_2$ can be appreciated at higher energy than the SM, while in 3~L-MoTe$_2$ the opposite case takes place. Experimental studies of the low-frequency modes in TMDs confirmed that the energies of the rigid oscillation modes can be well described within a linear chain model. The model considers every layer as a single point mass connected to the nearest neighboring layers with springs, see Fig. \ref{fig:196}(b). The interlayer interaction can be described by effective interlayer force constants $K_\text{j}$ (j = z for BM or j = x for SM). It is also assumed that the force constant does not depend on the number of layers. Finally, no interaction with the substrate is taken into account in the model. The evolution of the interlayer mode energies (expressed in cm$^{-1}$) as a function of the number of layers, $N$, is given within the model by \cite{lin2019ultralow}:

\begin{equation}\label{eq:LF}
\omega_{\text{j},N} (\alpha) =  \sqrt{\frac{K_{\text{j}}}{2\mu\pi^{2}c^{2}}\left ( 1-\cos\frac{(\alpha-1)\pi}{N}\right )}   ,
\end{equation}

\noindent where $K_\text{j}$ is the respective force constant, $\mu = 2 m_{\text{Te}} + m_{\text{Mo}}$ is the mass per unit area, $\alpha=2,3,4,\dots,N$ ($\alpha=1$ corresponds to the acoustic mode). The Raman-active SMs (BMs), observed in our experiment belong to the highest-energy (lowest-energy) branches of the corresponding set ($\alpha=2$) and will be hereafter referred to as $\omega_{j,2}$ and $\omega_{j,3}$ for bi- and tri-layer structure, respectively. Assuming that the energy of the BM (SM) in 2~L- and 3~L-MoTe$_2$ in the region A equals correspondingly  $\omega_{z,2}=27.2$~cm$^{-1}$ and $\omega_{z,3}=19.0~cm^{-1}$ ($\omega_{x,2}=19.3$~cm$^{-1}$ and $\omega_{x,3}=23.3$~cm$^{-1}$), the derived force constant value is $K_\text{z}=7.4 \times10^{19} \frac{\text{N}}{\text{m}^3}$ ($K_\text{x}=3.6 \times10^{19} \frac{\text{N}}{\text{m}^3}$). These force constants correspond to our previous results \cite{grzeszczyk2016raman} and other published data for MoTe$_2$ ($7.7 \times10^{19} \frac{\text{N}}{\text{m}^3}$ and $3.6 \times10^{19} \frac{\text{N}}{\text{m}^3}$, respectively) \cite{froehlicher2015unified}.

\begin{table*}
\caption{\label{tab:table}Energies of the observed interlayer vibrations excited with E$_\text{L}$=1.96~eV.}
\begin{ruledtabular}
\begin{tabular}{cccccccccc}
   & &\multicolumn{2}{c}{A}&\multicolumn{2}{c}{B}&\multicolumn{2}{c}{C}&\multicolumn{2}{c}{D}\\ 
 
 & & Mode & Energy &  Mode & Energy &  Mode & Energy &  Mode & Energy     \\ 
  & & & (cm$^{-1}$) &   & (cm$^{-1}$) &    & (cm$^{-1}$) &    & (cm$^{-1}$)     \\ \hline
 \multirow{4}{*}{2 L} & \multirow{3}{*}{BM} & &  & & & $\omega^{-}_{\text{z},2}$ & 23.6 & $\omega^{-}_{\text{z},2}$ & 23.6 \\
& & $\omega_{\text{z},2}$ & 27.2 & $\omega'_{\text{z},2}$ & 26.65 & $\omega'_{\text{z},2}$ & 28.0 & $\omega'_{\text{z},2}$ & 28.3  \\
& &   &   &   &   & $\omega^{+}_{\text{z},2}$ & 33.1 & $\omega^{+}_{\text{z},2}$ & 32.4  \\ 
 & SM &  $\omega_{\text{x},2}$ & 19.25 & $\omega'_{\text{x},2}$ & 19.1 & $\omega'_{\text{x},2}$ & 19.1 & $\omega'_{\text{x},2}$ & 19.0  \\
 \hline
 \multirow{3}{*}{3 L} & \multirow{2}{*}{BM} & &  & & & $\omega^{-}_{\text{z},3}$ & 15.35 & $\omega^{-}_{\text{z},3}$ & 15.2 \\
& & $\omega_{\text{z},3}$ & 19.0 & $\omega'_{\text{z},3}$ & 18.4 & $\omega'_{\text{z},3}$ & 19.25 & $\omega'_{\text{z},3}$ & 19.6  \\
 & SM &  $\omega_{\text{x},3}$ & 23.3 & $\omega'_{\text{x},3}$ & 23.4 & $\omega'_{\text{x},3}$ & 23.4 & $\omega'_{\text{x},3}$ & 23.4  \\
\end{tabular}
\end{ruledtabular}
\end{table*}

The spectra substantially change, when the layers are placed on the h-BN/SiO$_2$/Si substrate (region D). The SMs in 2~L- and 3~L-MoTe$_2$ are not affected by the presence of the h-BN substrate (see Fig. \ref{fig:196}(c)). On the contrary three (two) peaks can be observed in the spectrum measured on the D region as compared to one BM in 2~L- (3~L-) MoTe$_2$/SiO$_2$/Si. Dominant in the 2~L-MoTe$_2$/h-BN/SiO$_2$/Si heterostructure is the feature at $\omega'_{\text{z},2}=28.3$~cm$^{-1}$, at the energy slightly higher than the energy of the corresponding BM in the MoTe$_2$/SiO$_2$/Si ($\omega'_{\text{z},2}=27.2$~cm$^{-1}$). The main peak is accompanied by two other peaks $\omega^{+}_{\text{z},2}=32.4$~cm$^{-1}$ and $\omega^{-}_{\text{z},2}=23.6$~cm$^{-1}$, which emerge at its higher and lower energy sides. Two peaks in the spectrum of 3~L-MoTe$_2$/h-BN/SiO$_2$/Si heterostructure can be observed at the energies $\omega^{-}_{\text{z},3}=15.2$~cm$^{-1}$ and $\omega'_{\text{z},3}=19.6$~cm$^{-1}$. 

\begin{figure}[h!]
    \centering
    \includegraphics[width=\linewidth]{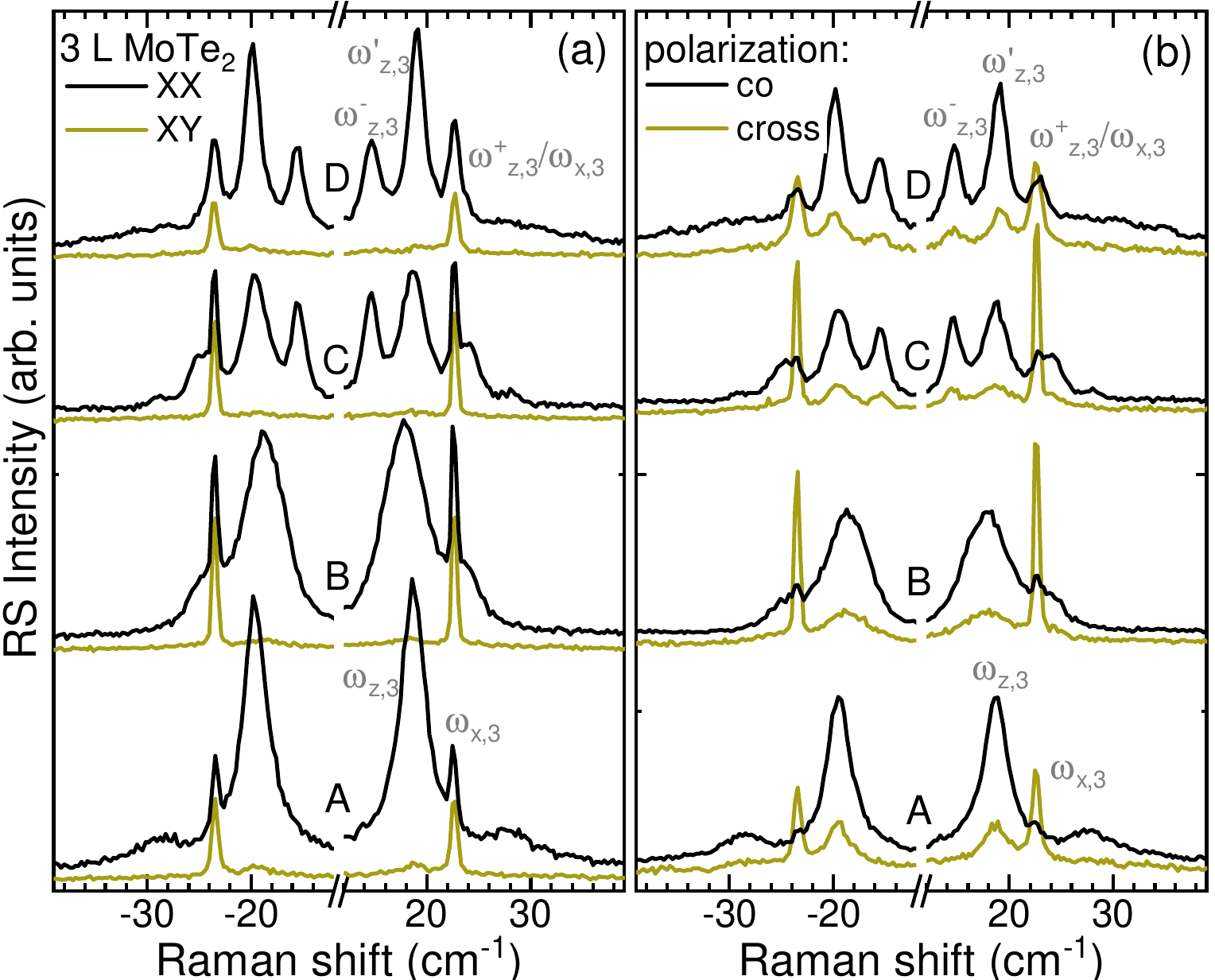}
    \caption{(a) Linearly polarized  and (b) helicity-resolved Raman scattering (E$_\text{L}$=1.96~eV) measurements of low-frequency Raman scattering for 3~L in the regions MoTe$_2$/SiO$_2$/Si (A), h-BN/MoTe$_2$/SiO$_2$/Si (B), h-BN/MoTe$_2$/h-BN/SiO$_2$/Si (C), and MoTe$_2$/h-BN/SiO$_2$/Si (D).}
    \label{fig:polarised}
\end{figure}

In order to attribute those peaks, polarization-sensitive measurements were performed (see Fig. \ref{fig:polarised}). It is well known, that the BMs have the A--symmetry and the SMs exhibit the E--symmetry \cite{loudon1964raman}. Therefore, in the backscattering geometry, the BM should not be observed in the (XY) configuration, in which the linear polarization of the scattered light is perpendicular to the linear polarization of the illuminating light, while both the BM and SM can be observed in the co-polarized (XX) configuration. As it can be appreciated in Fig. \ref{fig:polarised}(a), only SM modes are seen in the (XY) configuration from every region of 3~L MoTe$_2$ structure, while all the peaks can be observed in the (XX) configuration. This approach is most commonly used for characterizing symmetry of observed phonon modes. However, the presence of \mbox{E--symmetry} modes for both the co- and cross-polarization configurations may leave some uncertainties in resolving Raman spectra. Thus, in order to fully identify the described peaks, the same measurements have been carried out with circular polarization of light. The helicity-resolved Raman scattering can be a useful tool to distinguish contributions from degenerated modes of different symmetries. In this case, the selection rules ensure that A--symmetry phonons are only visible when helicities of incident and scattered light are the same (co-circular), while opposite situation (cross-circular) takes place for E--symmetry modes. \cite{chen2015helicity} This prediction is consistent with our results, presented in Fig. \ref{fig:polarised}(b). For region A, the co-circular configuration results in the spectrum dominated by one strong peak, which is assigned as BM ($\omega_{\text{z},3}$). In the corresponding cross-circular configuration, an additional peak becomes visible, which we assign as SM ($\omega_{\text{x},3}$). The SM peak can be clearly distinguished in regions C and D for both polarizations, which leads us to assume that we are dealing with two degenerate peaks of different symmetries. The fact that some modes are fairly visible in both polarizations is, in our opinion, the result of the used resonant excitation. It should be noted, however, that the intensity of peaks of a certain symmetry in individual measurements is significantly different. The polarization-resolved results for 3~L MoTe$_2$ structure confirm the attribution of the $\omega^{-}_{\text{z},3}$ and $\omega'_{\text{z},3}$ modes to the out-of-plane (BM) vibrations, as well as the degeneracy of the third peak, which includes contributions from both $\omega^{+}_{\text{z},3}$ and $\omega_{\text{x},3}$.

No qualitative difference between the spectra measured on the C and D regions of both 2~L and 3~L heterostructures can be noticed in Fig. \ref{fig:196}(c). This suggests that the effect of covering the 2~L and 3~L with an h-BN layer is noticeably weaker as compared to the significant influence of the bottom h-BN flakes. In our opinion, this is related to the fabrication process (see SM for details). Consequently, it may be expected that the results obtained on the B regions of the studied heterostructures may resemble those measured on the A-region due to the negligible interaction with the top h-BN layer. Surprisingly, in higher energies, an additional peak can be observed for the B region of the 2L sample. This may be attributed to the non-homogenous interaction between the MoTe$_2$ layer and the top h-BN flake as well as the strain effects due to the roughness of the SiO$_2$/Si substrate (see SM for details). The SM-related features in the B-region spectrum are rather similar to those observed in other (A, C, and D) regions as far as both the energy and the broadening are concerned. Importantly, for both structures, the energy of the main BM-related peaks in the B-region spectra for both 2~L and 3~L structures are slightly lower than those observed in the A region (as well as C and D regions).


It is generally accepted that the interaction of TMDs with the substrate can be neglected \cite{lee2015resonant} and slight differences between the low-energy Raman spectra of supported and suspended structures are typically ascribed to strain rather than to the interaction with the substrate \cite{o2017raman}. However, there are also reports on a substrate-induced mode in the low-frequency spectrum of Bi$_2$Te$_3$ nanoplates \cite{zhao2014interlayer} and a Raman mode splitting in few-layer black phosphorus encapsulated in h-BN \cite{urban2017observation}.

Let us consider an interaction between the bottommost MoTe$_2$ layer and the substrate, described by the a well-defined force constant  $K_\text{i}$. For a bilayer, there are two frequency branches $\omega'_{\text{z},2}$ and $\omega''_{\text{z},2}$ corresponding to the BMs, with energies given by:

\begin{subequations}
\label{eq:omega2}
\begin{equation}
\omega'_{\text{z},2} =  \sqrt{\frac{2K_{\text{z}}+
K_{\text{i}}+ \sqrt{4K_{\text{z}}^2+K_{\text{i}}^2}}{8\mu\pi^{2}c^{2}}}
\end{equation}
\begin{equation}
\omega''_{\text{z},2} =  \sqrt{\frac{2K_{\text{z}}+
K_{\text{i}}- \sqrt{4K_{\text{z}}^2+K_{\text{i}}^2}}{8\mu\pi^{2}c^{2}}} 
\end{equation}
\end{subequations}

\begin{figure}[h!]
    \centering
    \includegraphics[width=\linewidth]{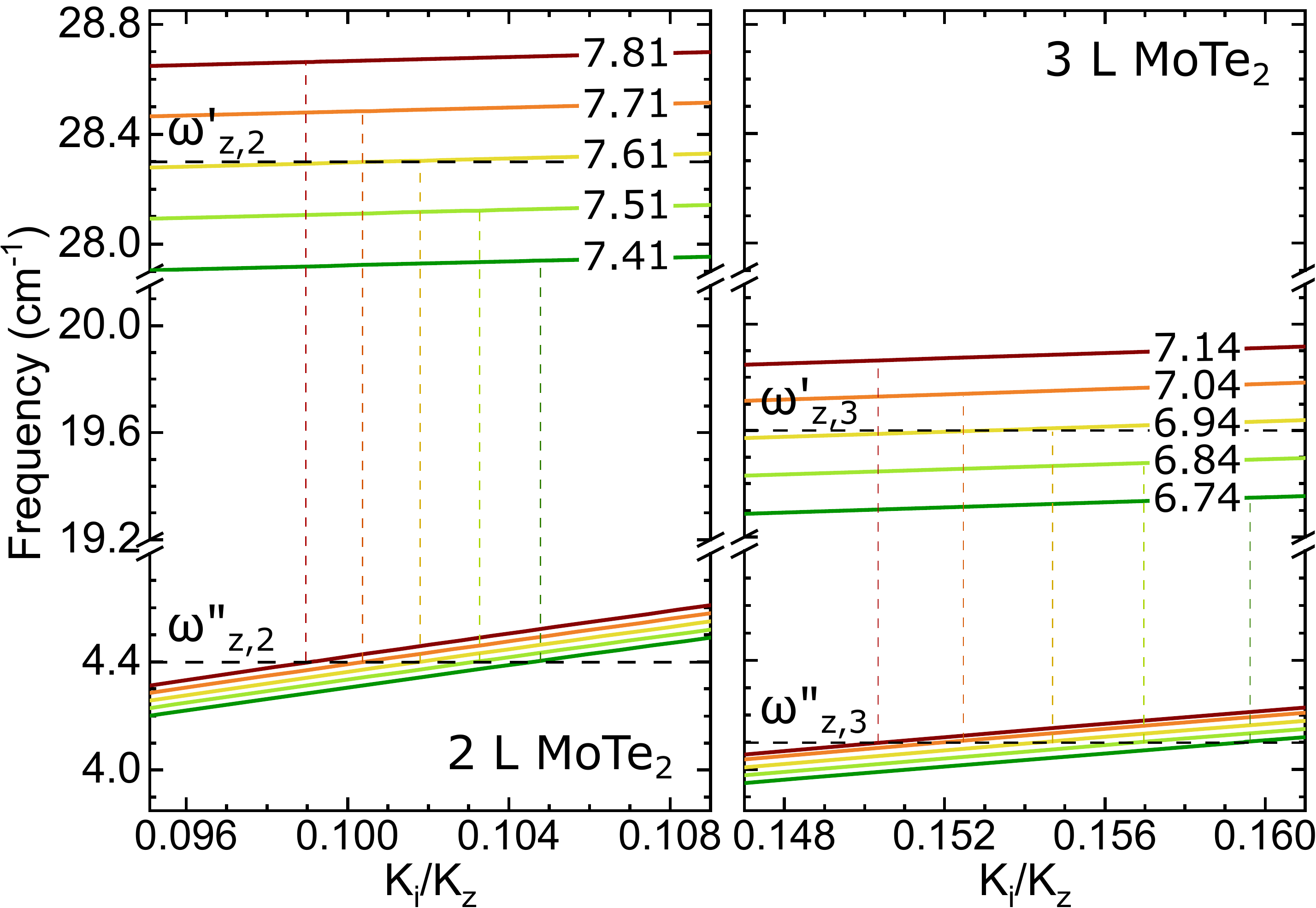}
    \caption{Theoretical evolution of vibrational modes in 2~L and 3~L MoTe$_2$ as a function of
$K_\text{i}$/$K_\text{z}$ - the ratio of the interface to the interlayer force constants for several values of $K_\text{z}$ (in $10^{19}\frac{\text{N}}{\text{m}^3}$ units).}
    \label{fig:omega}
\end{figure}

\noindent If $K_\text{i}$ equals zero (no interaction with the substrate), the modes correspond to the BM and the acoustic mode, of which only the former one is Raman-active and can be observed in the spectrum. On the other hand, if $K_\text{i}\neq0$ both modes become Raman active. Theoretical evolution of $\omega'_{\text{z},2}$ and $\omega''_{\text{z},2}$ with $K_\text{i}$/$K_\text{z}$ ratio for several values of force constants $K_\text{z}$ is shown in Fig. \ref{fig:omega}.   It can be seen that the $\omega'_{\text{z},2}$ energy only weakly depends on $K_\text{i}$. The evolution of the lower,  $\omega''_{\text{z},2}$ energy with the $K_\text{i}$/$K_\text{z}$ is much faster, however the energy is rather low for reasonable values of the $K_\text{i}$/$K_\text{z}$ ratio. The corresponding peak could be therefore hardly observed in the low-frequency Raman spectrum, as opposite to the reported results for Bi$_2$Te$_3$ and Bi$_2$Se$_3$ crystals \cite{zhao2014interlayer}. In our opinion, the $\omega''_{\text{z},2}$ vibrational mode contributes to the combination: $\omega^{+}_{\text{z},2}= \omega'_{\text{z},2}+\omega''_{\text{z},2}$ and $\omega^{-}_{\text{z},2}= \omega'_{\text{z},2}-\omega''_{\text{z},2}$ modes, which are present in the spectra of the C and D regions of the structure. Assuming $\omega''_{\text{z},2}=\frac{1}{2}(\omega^{+}_{\text{z},2}-\omega^{-}_{\text{z},2})=4.4$~cm$^{-1}$ (an average energy difference between the main BM peak and its satellites) one can get the corresponding $K_\text{i}$/$K_\text{z}=0.102\pm0.002$. As it can be appreciated in Fig. \ref{fig:omega}, the energy of the main peak $\omega'_{\text{z},2}=28.3$~cm$^{-1}$, which corresponds to the $K_\text{i}$/$K_\text{z}$ value stated above, can be reproduced with $K_\text{z}=(7.61\pm 0.05) \times10^{19}  \frac{\text{N}}{\text{m}^3}$. This value corresponds well to the force constant found for the A region of the 2~L MoTe$_2$. The interaction with the substrate can be therefore described with $K_\text{i}=(7.75\pm0.004) \times10^{18} \frac{\text{N}}{\text{m}^3}$.

Before the same model will be applied to the 3~L-MoTe$_2$ heterostructure, it should be noticed that just two BMs can be observed in the corresponding Raman scattering spectrum, while excited with E$_\text{L}$=1.96~eV light, as opposite to three peaks in the spectrum of similar 2~L-MoTe$_2$ structure. In fact, if the $\omega^{-}_{\text{z},3}$ peak is related to the difference: $\omega'_{\text{z},3}-\omega''_{\text{z},3}$, the $\omega^{+}_{\text{z},3}$ peak related to the sum $\omega'_{\text{z},3}+\omega''_{\text{z},3}$ should be present in the spectrum. The latter mode cannot be distinguished in the E$_\text{L}$=1.96~eV excited Raman spectrum, as it coincides with the SM in the structure. As was previously stated, the helicity-resolved measurments confirmed the degeneracy of this two phonons (see Fig. \ref{fig:polarised}(b)). This conclusion supports our model with the combined sum and differential modes, and justifies its application to the 3~L-MoTe$_2$ structure. The corresponding equation of motion cannot be solved analytically in that case, so the results of numerical analysis are presented in Fig. \ref{fig:omega}. Assuming $\omega''_{\text{z},3}=\frac{1}{2}(\omega^{+}_{\text{z},3}-\omega^{-}_{\text{z},3})=4.1$~cm$^{-1}$ one can get the corresponding $K_\text{i}$/$K_\text{z}=0.154\pm0.004$~cm$^{-1}$. Finally $K_\text{z}=(6.94\pm0.02) \times10^{19} \frac{\text{N}}{\text{m}^3}$ and $K_\text{i}=(1.07\pm0.02) \times10^{19} \frac{\text{N}}{\text{m}^3}$ can be found in order to reproduce the energy of the main peak $\omega'_{\text{z},3}=19.6$~cm$^{-1}$.

We propose the following scenario to explain our results. The bottom MoTe$_2$ interacts with the h-BN substrate in 2~L- and 3~L-MoTe$_2$. Because of the high uniformity of the substrate (see SM for the AFM topography), the resulting vibrational mode has a well-defined vibration frequency and contributes to the combination of Raman modes, as can be seen in the spectra. Therefore, crucial for the effect of encapsulation is the h-BN substrate, which provides a flat support for the TMD layer \cite{jadczak2019room,kang2019enhancing}. This minimizes the effect of strain and disorder resulting from the roughness of the SiO$_2$/Si substrate. The resemblance of the spectra from regions C and D suggests that the top \mbox{h-BN} layer does not strongly interact with the MoTe$_2$ layers but rather provides protection from ambient conditions \cite{son2017graphene, lee2015highly}. The results from region B can also be addressed within our model. Characteristic of both the 2~L- and 3~L-MoTe$_2$ case is the lowering of the main BM peak energy as compared to the uncovered structures. The energy redshifts suggest the presence of substantial strain in the structure. As schematically shown in Fig. \ref{fig:196}(a), the region B is not uniform and the deformation can change from point to point.

Our model does not address all features of the presented results. The force constants, which could be used to reproduce the observed Raman spectra seem to depend on the number of layers, however, the experimental uncertainty can be partially responsible for the effect. It is important to mention that similar results devoted to the interaction effects between a substrate and a TMD layer were reported for WS$_2$/h-BN structures. \cite{lin2019cross} The authors shown that the reduced symmetry of heterostructures leads to the activation of new interlayer phonons in the Raman spectra. Those modes could be observed due to their intensity enhancement resulting from the resonance of the excitation with the specific excitonic transition in WS$_2$. 


In conclusion, the effect of h-BN encapsulation on the low-frequency Raman scattering in 2~L- and 3~L-MoTe$_2$ has been investigated. Three BMs have been observed in the spectra of few-layer MoTe$_2$ deposited on/encapsulated in h-BN as compared to one mode for the flake deposited on SiO$_2$/Si. The additional satellite peaks have been assigned to the two-phonon scattering effect. The SM does not seem to be affected by changing the substrate. The results have been analyzed in terms of a linear chain model. The observation has been explained by introducing to the model a well-defined interaction between the lowest MoTe$_2$ layer and the h-BN substrate. The interaction slightly affects the energy of the main BM and activates an otherwise inactive, zero-frequency acoustic mode. The latter mode cannot be directly observed in the spectrum but it contributes to the combination modes in a MoTe$_2$ structure. 

\section*{Supplementary Material}
See supplementary material for experimental details, AFM analysis, Raman spectra in non-resonant regime and data from other samples.

\begin{acknowledgments}
The work has been supported by the National Science Center, Poland (grant no. 2017/27/B/ST3/00205, 2017/27/N/ST3/01612, 2017/24/C/ST3/00119), the ATOMOPTO project (TEAM programme of the Foundation for Polish Science co-financed by the EU within the ERDFund), EU Graphene Flagship project (ID: 785219) and Ministry of Education, Youth and Sports of the Czech Republic under the project CEITEC 2020 (LQ1601).
\end{acknowledgments}

\section*{AIP Publishing Data sharing policy}
The data that support the findings of this study are available from the corresponding author upon reasonable request. \\

The following article has been accepted by Applied Physics Letters. After it is published, it will be found at {\color{blue}\href{https://dx.doi.org/10.1063/1.5128048}{https://dx.doi.org/10.1063/1.5128048}}

\bibliography{aipsamp}

\newpage
\onecolumngrid
\setcounter{figure}{0}
\setcounter{section}{0}
\renewcommand{\thefigure}{S\arabic{figure}}
\renewcommand{\thesection}{S\arabic{section}}

\begin{flushleft}

	\textbf{\textsf{{\Large{Supplementary Material: \\ Breathing modes in few-layer MoTe$_2$ activated by h-BN encapsulation}}}}
	\vskip0.5\baselineskip{\textsf{\hspace{12mm}M.~Grzeszczyk,{$^{1}$} M.~R.~Molas,{$^{1}$} M.~Barto\v{s},{$^{2,3}$} K. Nogajewski,{$^{1}$} M. Potemski,{$^{1,2}$} A.~Babiński{$^{1}$}}}\\
	\small{{\em
	\hspace{12mm}$^{1)}$Institute of Experimental Physics, Faculty of Physics, University of Warsaw, ul.
Pasteura 5, 02-093 Warsaw,\\ 
\hspace{12mm}Poland \\
	\hspace{12mm}$^{2)}$Laboratoire National des Champs Magn\'etiques Intenses, CNRS-UGA-UPS-INSA-EMFL, 25, avenue des Martyrs,\\ \hspace{12mm}38042 Grenoble, France \\ 		
	\hspace{12mm}$^{3)}$Central European Institute of Technology, Brno University of Technology,  Purky\v{n}ova 656/123, 612 00 Brno, Czech\\ \hspace{12mm}Republic}}

\end{flushleft}

\renewcommand{\thesection}{S\arabic{section}}
\renewcommand{\thefigure}{S\arabic{figure}}


\onecolumngrid
\section{Sample fabrication}\label{sec:fab}

\begin{figure}[h!]
    \centering
    \includegraphics[width=.7\linewidth]{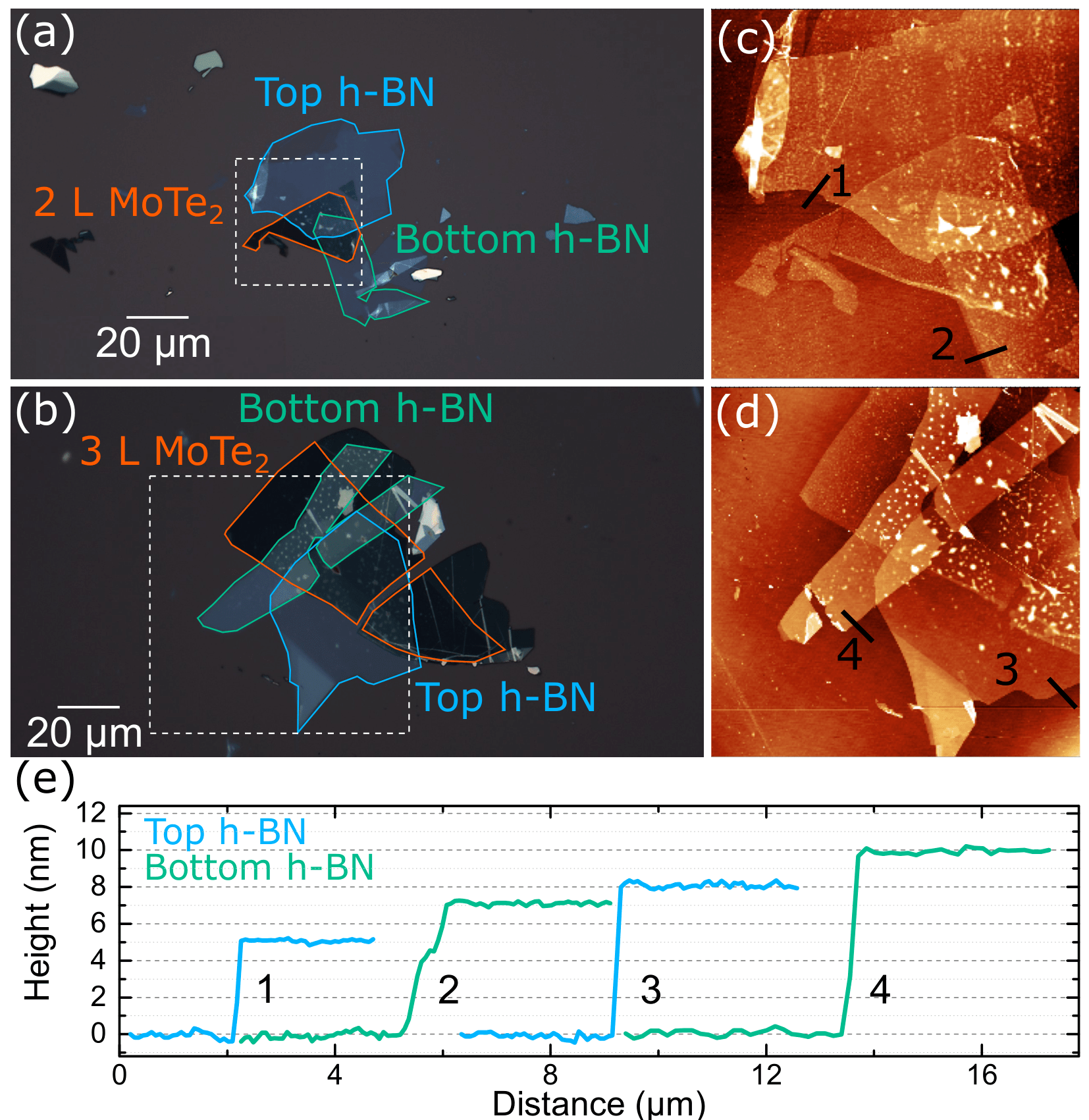}
    \caption{(a)-(b) Optical microscope images and (c)-(d) atomic force microscope topography of investigated 2~L and 3~L MoTe$_2$ heterostructures. (e) Height profiles along straight lines marked in (c)-(d).}
    \label{fig:samples}
\end{figure}

Investigated samples comprised of thin MoTe$_2$ and h-BN layers fabricated by two-stage PDMS-based mechanical exfoliation of bulk crystals purchased from HQGraphene. As a substrate, a SiO$_2$ (90~nm)/Si wafer was used. In order to ensure the best quality of the substrate surface, they were first cleaned (placed in acetone in the ultrasonic washer then rinsed in isopropanol and deionized water) and annealed at 200$^{\circ}$C. They were kept on hot-plate until the first non-deterministic transfer of h-BN flakes. This procedure minimizes the occurrence of air pockets and ensures good adhesion between the flake and the substrate. It should be underlined that subsequent layers were transferred deterministically, which adds the possibility of additional contaminants on the interfaces between the constituent layers. In an attempt to reduce inhomogeneities between each transfer, the sample was annealed. The 2~L- and 3~L-MoTe$_2$ flakes were placed on the surface partially on top of the SiO$_2$ (90~nm)/Si substrate and partially on the h-BN layer. Such heterostructures were partially capped with top h-BN flakes. The complete structures were annealed at 120$^{\circ}$C for 1.5 hour in order to ensure the best layer-to-layer and layer-to-substrate adhesion and to eliminate a substantial portion of air pockets at interfaces between the constituent layers.

Fig. \ref{fig:samples}(a)-(b). presents an optical microscope image of the investigated heterostructures. The sample topography, as measured by atomic force microscopy, is shown in Fig. \ref{fig:samples}(c)-(d). Height profile measurements (performed along straight lines marked in (c)-(d)), which were performed in order to unambiguously determine the thickness of layers, are presented in Fig. \ref{fig:samples}(e). The top h-BN layer was 5~nm (8~nm) thick and the bottom h-BN layer was 7~nm (10~nm) thick in the 2~L- (3~L) case.

\section{Surface roughnesses of substrates}\label{sec:roughness}

\begin{figure}[h!]
    \centering
    \includegraphics[width=.9\linewidth]{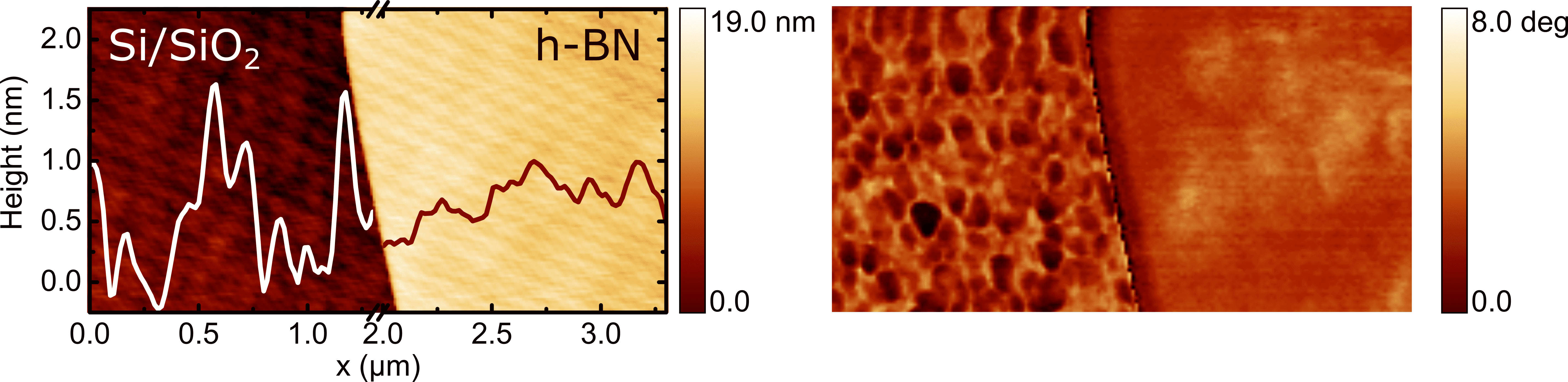}
    \caption{(left) Topography image of h-BN on SiO$_2$ with height line profile and (right) phase shift signal measured in th intermittent-contact AFM.}
    \label{fig:afm}
\end{figure}

In the left-hand side panel of Fig. \ref{fig:afm} the AFM image of an h-BN flake non-deterministically transferred on Si/SiO$_2$ substrate is shown. The Figure illustrates the effect of substrate on the studied material. To quantify the roughness of both surfaces, we show the corresponding height profiles measured on the hBN flakes and the SiO$_2$ substrate. As it can be seen in the Figure, the surface roughness of the Si/SiO$_2$ substrate is significantly larger as compared to the h-BN flake, which can be seen as much larger variation of heights of the Si/SiO$_2$ (almost 2 nm) as compared to the h-BN (about 0.5 nm). This confirms that h-BN thick layers can work as atomically flat substrates, which is consistent with previous reports. \cite{xue2011scanning,quereda2014single}  In the right-hand side panel of Fig. \ref{fig:afm} the corresponding phase image is shown. That powerful extension of AFM tapping mode is very sensitive to local variations in the material properties \cite{boussu2005roughness,magonov1997phase} and it is often used for its superior contrast of nanoscale features. The presented results demonstrate the difference in topography of Si/SiO$_2$ and h-BN surfaces. The strong contrast variation clearly reveals the non-uniform nature of the SiO$_2$ surface.

\section{Experimental setup}

Raman scattering measurements were carried out in the backscattering geometry using 1.96~eV excitation from a He-Ne laser and 2.41~eV excitation from an Ar$^+$ ion laser. The excitation light was focused by means of a 100x magnification long working distance objective. The spot diameter of the focused beam was equal to about 1 $\upmu$m. The excitation power focused on the sample was kept at 80 $\upmu$W during all measurements to avoid local heating. The scattered light was collected via the same microscope objective, sent through a 1~m monochromator and detected with a liquid-nitrogen cooled CCD camera. To detect low-energy Raman scattering up to about $\pm$10~cm$^{-1}$ from the laser line, a set of Bragg filters were implemented in both excitation and detection paths. The linear polarization of the measured signal was analyzed using a set of polarizers and a half-waveplate, while for helicity resolved measurements additionally a quarter-wave plate was implemented.

\section{Supplementary Raman spectra of MoTe$_2$ flakes.}

\subsection{Non-resonant Raman scattering}

\begin{figure}[h!]
    \centering
    \includegraphics[width=.9\linewidth]{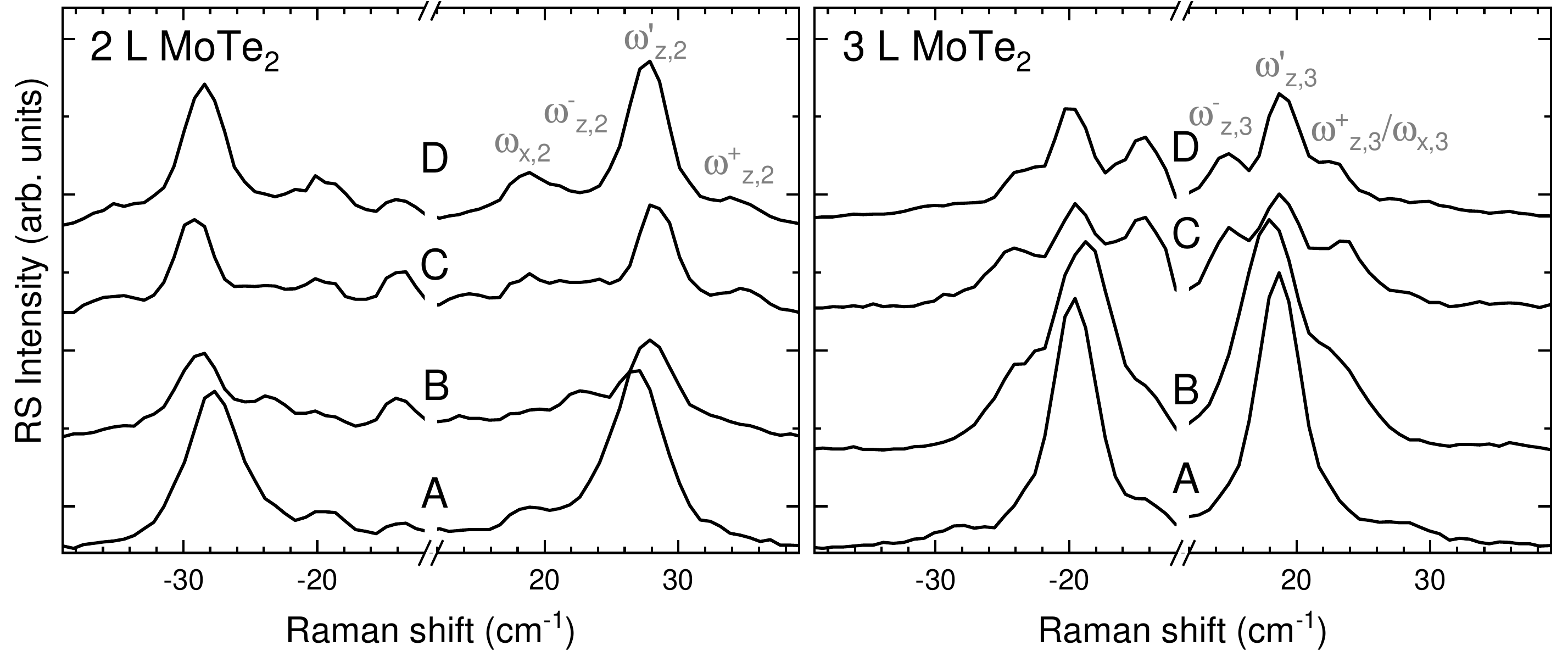}
    \caption{Low-frequency Raman spectra (E$_\text{L}$=2.41~eV) of 2~L and 3~L MoTe$_2$ in MoTe$_2$/SiO$_2$/Si (A), h-BN/MoTe$_2$/SiO$_2$/Si (B), h-BN/MoTe$_2$/h-BN/SiO$_2$/Si (C), and MoTe$_2$/h-BN/SiO$_2$/Si (D).}
  \label{fig:241}
\end{figure}

The overall lineshape of the low-frequency Raman spectrum strongly depends on the excitation energy, which is related to resonance effects \cite{yang2017}. In particular, we previously observed, that the SM in few-layer MoTe$_2$ is very weak in the non-resonant Raman scattering spectrum. \cite{grzeszczyk2016} The spectra measured with E$_\text{L}$=2.41~eV excitation in this experiment are shown in Fig. \ref{fig:241}. The SMs can be hardly seen in the spectrum from the A region, while the expected $\omega^{+}_{\text{z},3}$ peak can be distinguished in the spectra from the C and D regions of the 3~L-MoTe$_2$ heterostructure. 

\subsection{MoTe$_2$ flakes covered with h-BN}

\begin{figure}[h!]
    \centering
    \includegraphics[width=.9\linewidth]{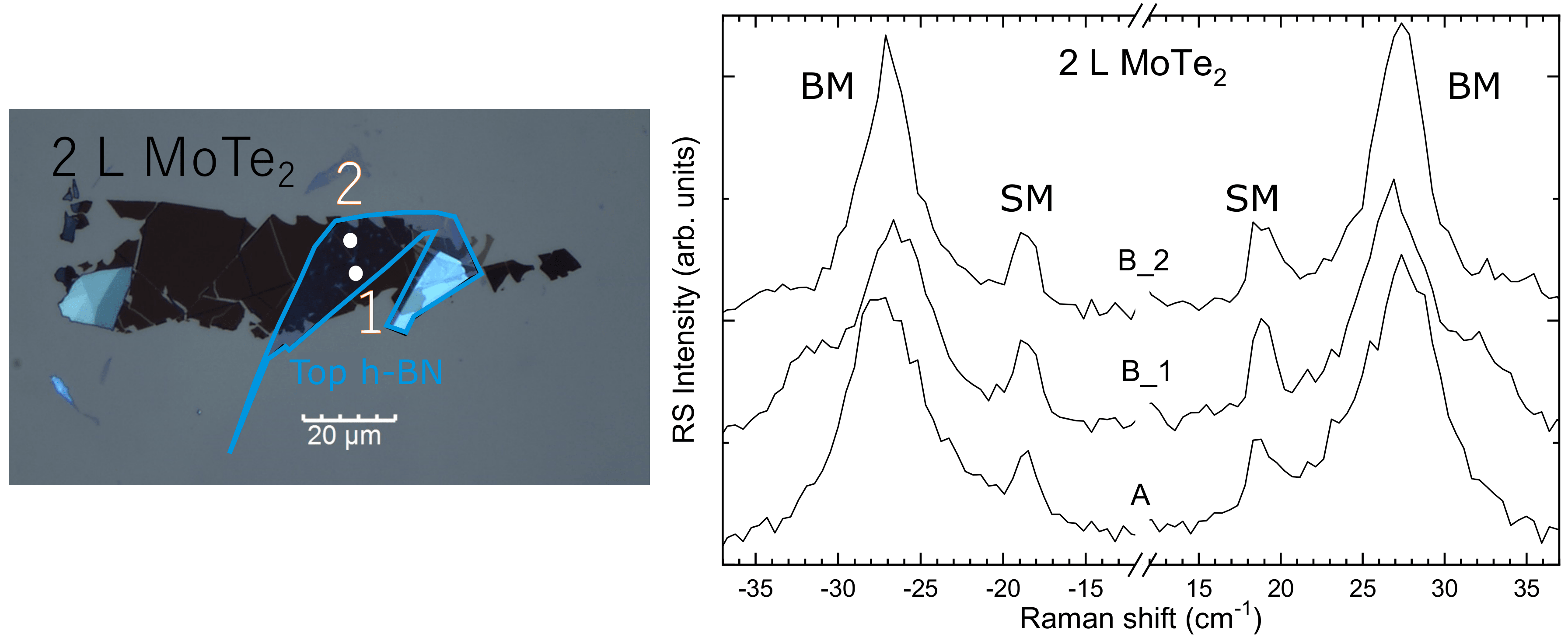}
    \caption{(left-hand side) Optical microscope image of the structure with marked spots, where the measurements were taken. (right-hand side) Low-frequency Raman spectra (E$_\text{L}$=1.96~eV) of 2~L MoTe$_2$ in MoTe$_2$/SiO$_2$/Si (A) and two different spots in h-BN/MoTe$_2$/SiO$_2$/Si (B).}
  \label{fig:cov}
\end{figure}

In an attempt to better understand the results observed in the B region, another sample with 2~L MoTe$_2$ covered with thin top h-BN flake was fabricated. As it is presented in Fig.~\ref{fig:cov}, one can easily find areas in which the additional peak is not visible due to the bigger size of the sample.

\subsection{Fully encapsulated samples}
\begin{figure}[h!]
    \centering
    \includegraphics[width=\linewidth]{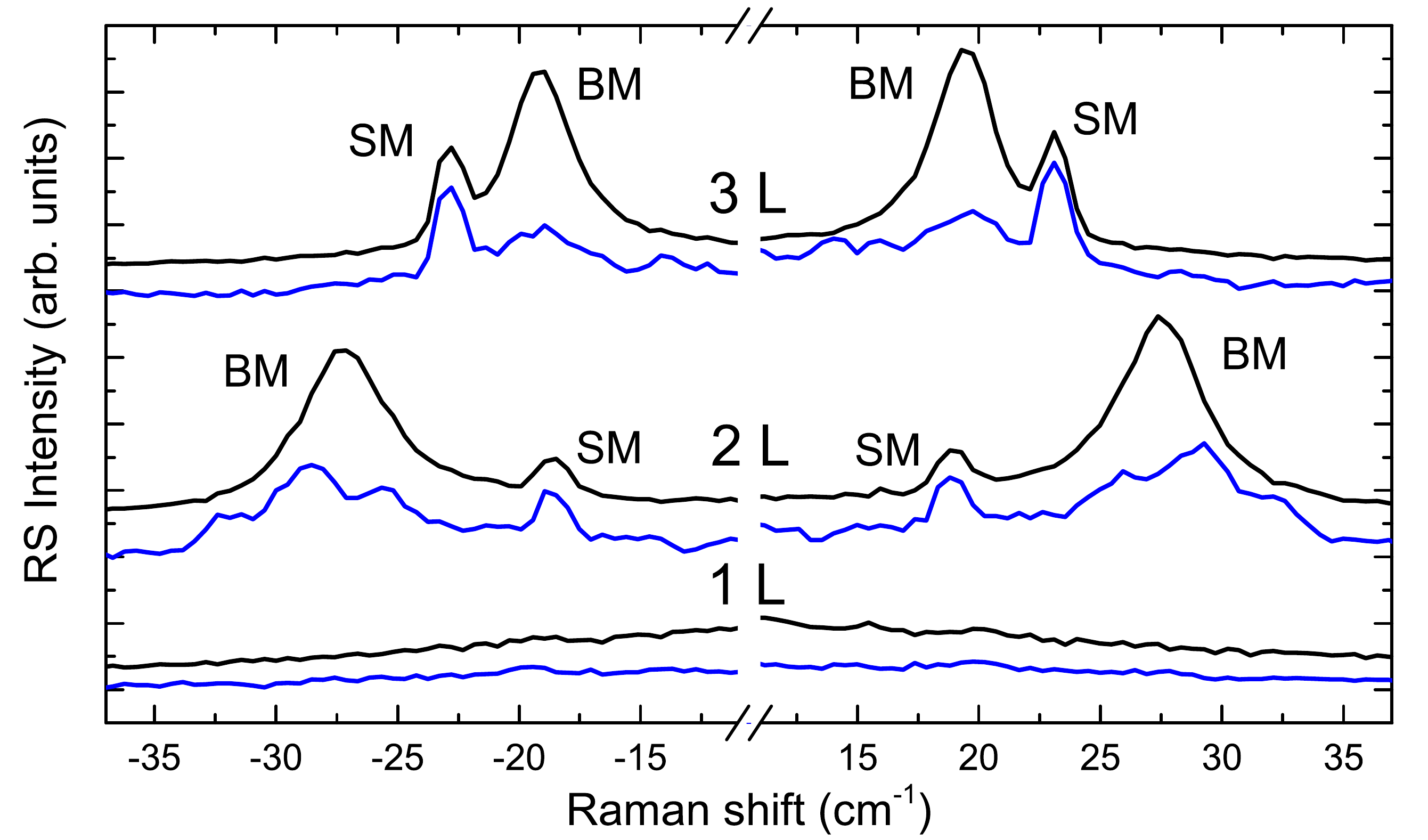}
    \caption{Low-energy Raman spectra (E$_\text{L}$=1.96~eV) of MoTe$_2$ thin layers (black curves) deposited on Si/SiO$_2$ substrate and (blue curves) encapsulated in h-BN flakes.}
    \label{fig:enc}
\end{figure}

The low-energy Raman spectra of MoTe$_2$ thin layers from a monolayer (1~L) to trilayer (3~L)  are shown in Fig.~\ref{fig:enc}. The spectra were measured on two different samples: as deposited on Si/SiO$_2$ substrate and encapsulated in h-BN flakes. Note that this sample was fabricated and investigated prior to the structures studied in the main text. To describe the presented Raman spectra, we can start the analysis with the results obtained for 1~L. As can be seen in the Figure, there are no observed low-energy mode is present in both structures. It is well known, that no rigid interlayer vibrations in the case of a monolayer are expected if the interaction between the monolayer and the substrate is neglected~\cite{lee2015,froehlicher2015,grzeszczyk2016}. As we discuss in the main text, the structure-substrate interaction leads to the activation of the acoustic mode, which consequently should be observed in the Raman spectrum. Our predictions suggest that its energy is of about 5~cm$^{-1}$ (see Fig. 3 in the main). However, we were not able to detect it as the used Bragg filter allows to measure a signal of around $\pm11$~cm$^{-1}$ from a laser line. The results obtained for 2~L and 3~L show that there is a significant effect of the h-BN flakes on their Raman spectra. The observed structures of breathing modes in h-BN encapsulated layers are very similar to the ones presented in Fig.~2 in the main text. Particularly, a triple structure of breathing modes can be observed straightforward for a bilayer encapsulated in h-BN. The thicknesses of the bottom h-BN flake for the studied encapsulated MoTe$_2$ layers shown in Fig.~\ref{fig:enc} differs from the that of the samples described in the main text ($\sim$100 nm versus $\sim$10 nm). This suggests that the thickness of h-BN flakes does not affect the apparent structure of breathing modes (three peaks). However, as it can be noted, their thickness may influence the measured intensity of phonon modes due to the occurred interference-effect.  

\section{Linewidth of the main breathing modes in Raman spectra}

\begin{figure}
    \centering
    \includegraphics[width=\linewidth]{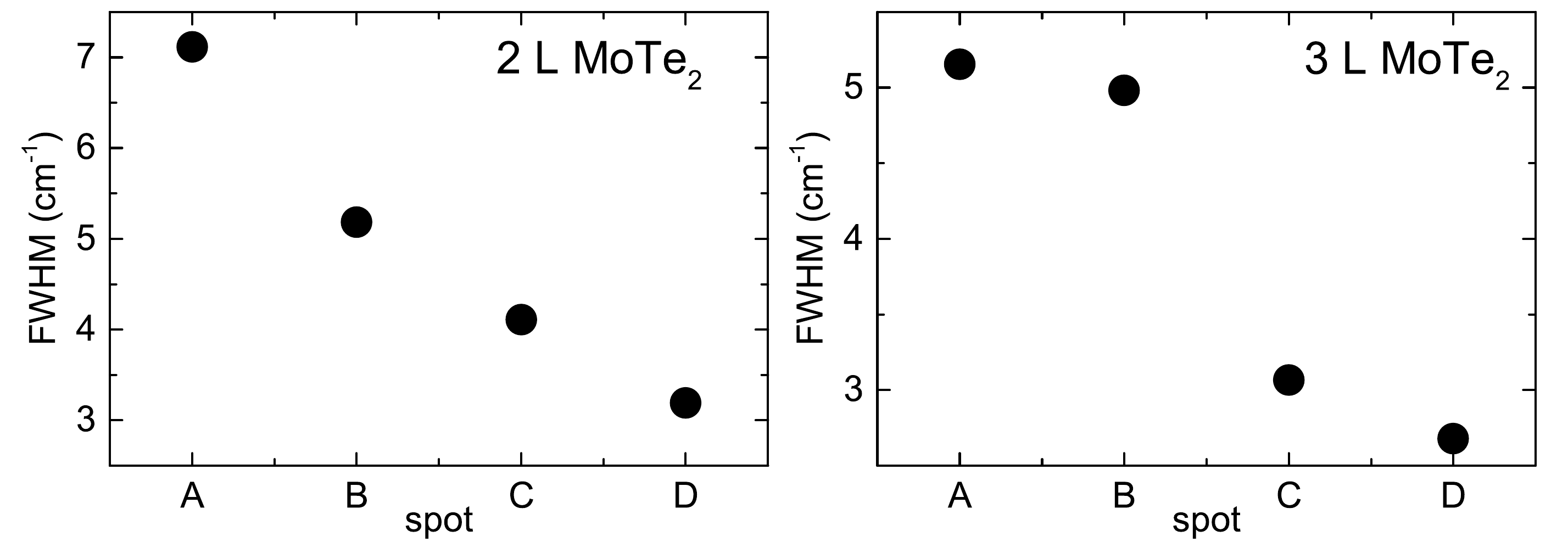}
    \caption{FWHM of main breathing mode $\omega_{z,2}$/$\omega'_{z,2}$ and $\omega_{z,3}$/$\omega'_{z,3}$ measured on different spots of the heterostructures.}
    \label{fig:fwhm}
\end{figure}

Fig. {\color{blue}S6} illustrates the change of the linewidth of main Raman modes $\omega_{z,2}$/$\omega'_{z,2}$ and $\omega_{z,3}$/$\omega'_{z,3}$ measured on the heterostructures. For both samples a significant decrease of the FWHM can be observed from spot A to spot D. For pristine MoTe$_2$ flake linewidth of the $\omega_{z,2}$ peak equals $\sim$7~cm$^{-1}$(5~cm$^{-1}$) for 2~L (3~L) thickness. After covering the flake with the h-BN layer this value is decreasing. This behavior can be understood in terms of surface quality. High roughness of the Si/SiO$_2$ substrate leads to the inhomogeneous broadening of the Raman peaks. As was previously shown in Section~\ref{sec:roughness}, h-BN flakes have much flatter surfaces. Capping MoTe$_2$ with the h-BN layer improves the quality of its surface, and as a result, the narrowing of the phonon peak can be observed. This effect is most evident in the C and D areas, where MoTe$_2$ flake is placed on the h-BN layer. In this case, atomically flat substrates lead to the peak's linewidth reduction by half. Our reasoning can explain the decrease of the linewidth between A, B, and C areas. However, we see a further drop in the peak's FWHM between C and D areas in both structures. This result is not surprising. As it was mentioned before h-BN can be considered as atomically flat substrates. The good adhesion between h-BN and MoTe$_2$ is also induced by our fabrication process. The interface between the top h-BN layer and MoTe$_2$ is of considerably lower quality. Our result reveals the presence of strain and many air-pockets in part of the samples with h-BN capping layer. These impurities and defects add to the broadening of the phonon linewidth.

\end{document}